\documentclass[fleqn,12pt,twoside]{article}
\usepackage{espcrc1}

\usepackage{graphicx}
\usepackage[figuresright]{rotating}
\usepackage{wrapfig}

\newcommand{\AmS}{{\protect\the\textfont2
  A\kern-.1667em\lower.5ex\hbox{M}\kern-.125emS}}

\title{A${(e,e'p)}$ reactions at GeV energies}

\author{Dimitri Debruyne and Jan Ryckebusch \thanks{e-mail : jan.ryckebusch@rug.ac.be}
        \address{Department of Subatomic and
        Radiation Physics, 
        Ghent University, \\ 
        Proeftuinstraat 86, B-9000 Gent, Belgium}}
       
\begin{document}

\maketitle

\begin{abstract}
An unfactorized and relativistic framework for calculating A$(e,e'p)$
observables at typical JLAB energies is presented.  Results of
$(e,e'p)$ model calculations for the target nuclei $^{12}$C
and $^{16}$O are presented and compared to  data from SLAC and
JLAB.
\end{abstract}

\section{Introduction and physics motivation}
Exclusive $A(e,e'p)$ reactions at sufficiently large values of the
four-momentum transfer $Q^2$ are an excellent tool to learn more about
different aspects of nuclear and nucleon structure in a regime where
one expects that both hadronic and partonic degrees-of-freedom
may play a role. Amongst the physics subjects which can be probed with
the aid of exclusive electro-induced proton knockout from nuclei we
mention the study of the short-range structure of nuclei.  Here, one
of the major questions to be addressed is whether there are any
hadronic components in the nucleus that carry large momenta.  
Another question which has attracted
the attention of the nuclear physics community for many years, is the
apparent robustness of nucleons in the medium.  Various techniques
have been used to determine to what extent nucleons are modified in
the medium.  One of the more recent and more promising ones are
double polarization $(\vec{e},e'\vec{p})$ processes from nuclei
\cite{sonja}.  These processes are meant to provide stringent tests of
constituent-quark models predicting measurable medium modifications of the 
electromagnetic form factors of bound nucleons \cite{ransome}.
Another subject of current investigation is the problem of nuclear
transparency in proton knockout from nuclei.  Amongst other things,
these investigations address the question whether there are any signatures
for the onset of non-hadronic degrees of freedom for a given
distance scale.  The latter can be varied by exploring ranges of
$Q^2$.

All of the above physics topics are awaiting further exploration at
facilities like Jefferson Laboratory (JLAB), the Mainz microtron
(MAMI) and the MIT-Bates electron accelerator, provided that one can
develop reliable models which provide a realistic description of the
final-state interactions (FSI) in A$(e,e'p)$ reactions.  Popular
frameworks to treat FSI effects in modelling ${A(e,e'p)}$ reactions
can roughly be divided in two major classes.  At ``low'' energies
$(p_p \leq 1$~GeV/c) most models use optical potentials to determine
the scattering wave function for the ejected proton.  The parameters
in the optical potentials are usually obtained from global fits to
elastic $p+A$ scattering data and are available in both relativistic
and non-relativistic forms.  At higher energies $(p_p \geq 1$~GeV/c),
on the other hand, the Glauber model offers good prospects to deal
with the distortions which the ejectile undergoes.  In the Glauber
model, the FSI is calculated directly from the elementary
nucleon-nucleon scattering amplitudes $p+N \longrightarrow p+N$.  One
of the major goals of the work presented here, is to provide a
framework in which the low and high-energy regime can be bridged.  In
particular, we deem that a ``smooth transition'' between the low and
high-energy regime is a prerequisite for any model that claims to
provide a realistic description of the final-state interactions (FSI).
The separation between a ``low'' and ``high-energy'' regime in
modelling FSI's, is usually done on the basis of the observation that
for lab proton momenta $p_p$ exceeding 1~GeV/c the elementary
proton-nucleon scattering process becomes highly inelastic.

\section{Relativistic Eikonal Model for A$\mathbf{(e,e'p)}$}

The model for the description of $A + e \longrightarrow (A-1) + e' +p$
processes which we developed has the following features.  First, it can
accommodate relativity in both the description of the electron-proton
coupling and the nuclear dynamics.  Further, it can be used in
combination with both the ``optical potential'' and the ``Glauber''
framework for dealing with the final-state interactions.  Furthermore,
the framework does not use any partial-wave expansion in its numerical
calculations and can therefore be used to calculate cross sections at
very high values of $Q^2$ in a rather elegant manner.  In our model,
A$(\vec{e},e'\vec{p})$ observables  
can be computed for any target nucleus with $A \ge 4$ provided that
its ground-state wave function can be reasonably well described within the
context of a relativistic mean-field model.  Finally, we provide a
completely unfactorized description of the reaction dynamics.  This
means that our model is \textbf{NOT} based on an expression of the
type 
\begin{equation}
\frac {d^6 \sigma} {d T_p d \Omega_p d \epsilon ' d \Omega _{\epsilon
'}} (e,e'p)  =
\frac {p_p E_p} {(2\pi)^3}  \sigma _{ep} 
P \left( \vec{p}_m, E_m \right) \; .
\end{equation}        
The use of this factorized expression was abandoned in the description
of low-energy $A(e,e'p)$ process many years ago, but still seems to be
common practice when it comes to modeling $A(e,e'p)$ processes at
higher energies.
In an unfactorized framework, the $A(\vec{e},e'p)$ cross section takes
on the well-known form in the one-photon exchange approximation
\begin{eqnarray}
& & \frac {d^5 \sigma}  {d \Omega _p d \epsilon ' d \Omega
_{\epsilon '}} (\vec{e},e'p)   =   
{1 \over 8 \pi^3 } \frac {M_p M_{A-1} p_p} {M_A} 
 f_{rec}^{-1} \sigma_{Mott}
\nonumber \\ 
& & \times
\Biggl[ v_T {W_T}
+ v_L {W_L}
+ v_{LT} {W_{LT}} \cos \phi _p
+ v_{TT} {W_{TT}} \cos 2 \phi _p
+ h \biggl[ v'_{LT} {W'_{LT}} + 
v'_{TT} {W'_{TT}} \biggr] \Biggr] \; ,
\label{eq:eepnn}
\end{eqnarray}
where $h$ denotes the electron helicity and the structure functions
$W$ depend on the variables (q,$\omega$,p$_p$,$\theta _p$). 
In the Impulse Approximation (IA) the
structure functions $W$ are determined by matrix elements of the type
\begin{equation}
m_F^{fi} \left(J ^{\mu}\right) = \left< K_f s_f \left| J^{\mu} \right|
K_i s_i \right> =
\overline{u} _f \Gamma ^{\mu} \left( K_f s_f, K_i  s_i \right) u_i \; ,
\end{equation}
where in the final state one has a scattering wave function
$u_f(K_f,s_f)$ and in the initial state a bound-state wave function
$u_i(K_i,s_i)$. Further,  $\Gamma ^{\mu} (K_f \; s_f, K_i \; s_i)$
represents the off-shell electron-proton coupling. 
In our model the relativistic bound-state wave functions are obtained
within the context of the ``$\sigma - \omega$ model'' which is a
relativistic quantum-field theory for nucleons ($\psi$) interacting
through pions ($\pi$'s), vector mesons ($\rho$'s and $\omega$'s) and
scalar mesons ($\sigma$'s) \cite{serot}. The model is usually solved
in the Hartree approximation by replacing the scalar and vector
meson-field operators by their expectation values.

The determination of the off-shell electron-proton coupling is amongst
the weakest parts of any relativistic approach to $A(e,e'p)$ processes
!  Indeed, there is some arbitrariness which translates itself in a
variety of recipes which all provide identical results for on-shell
particles but may lead to substantially different predictions when
off-shell nucleons are involved.  Amongst the more popular
forms for the electron-proton coupling  which have been used over the
years, we mention
\begin{eqnarray}
\label{eq:cc1}
\Gamma_{cc1}^{\mu} & = & G_{M} (Q^{2}) \gamma^{\mu} -
\frac{\kappa}{2M} F_{2} (Q^{2}) (K_{i}^{\mu}+K_{f}^{\mu}) \; , \\
\label{eq:cc2}
\Gamma_{cc2}^{\mu} & = & F_{1} (Q^{2}) \gamma^{\mu} + \imath
\frac{\kappa}{2M} F_{2} (Q^{2}) \sigma^{\mu\nu} q_{\nu} 
\; .
\end{eqnarray}

In determining the scattering wave functions we adopt the relativistic
eikonal approximation.  This method belongs to the class of
semi-classical approximations which become ``exact'' in the limit of
small de Broglie ($db$) wavelengths, $\lambda _{db} \ll a$, where
$a$ is the typical range of the potential in which the particle is
moving.  For a particle moving in a relativistic (optical) potential,
the scattering wave function has the following form in the eikonal
approximation 
\begin{equation}
\psi_{\vec{k},s}^{(+)} \sim 
\left[
\begin{array}{c}
1 \\
\frac{1}{E+M+{V_{s}-V_{v}}} \vec{\sigma} \cdot \vec{p} 
\end{array} 
\right]
e^{\imath \vec{p}_p \cdot \vec{r}} {e^{\imath S(\vec{r})}}
\chi_{\frac{1}{2}m_{s}} \; .
\label{eq:thescatteringwave}
\end{equation}
This wave function differs from a relativistic plane wave in two
respects.  First, there is a dynamical relativistic effect from the
scalar $V_s$ and vector $V_v$ potential, which enhances the
contribution from the 
lower components.  Second, the wave function contains an eikonal phase
which is determined by integrating the central ($V_c$) and spin-orbit
($V_{so}$) parts of the distorting potentials along the escaping particle's 
(asymptotic) trajectory ($ \vec{r} \equiv (\vec{b},z)$)
\begin{equation}
\imath S(\vec{b},z)  =  - \imath \frac{M}{K} \int_{-\infty}^{z} dz' \, \biggl[
{V_{c} (\vec{b},z')} 
+ {V_{so} (\vec{b},z')}   
[ \vec{\sigma} \cdot (\vec{b}
\times \vec{K} )- \imath Kz'] \biggr] \; ,
\label{eq:eikonalphase}
\end{equation}
where $\vec{K} \equiv \frac {1} {2} \left( \vec{p} _p + \vec{q} \right) $.
For proton lab momenta exceeding 1 GeV/c, the use of optical
potentials is no longer justifiable in view of the highly inelastic
character of the elementary processes.  In this energy regime, an
alternative description of FSI processes is provided in terms of the
Glauber multiple-scattering theory.  In such a framework the
A-body wave function in the final state reads 
\begin{equation}
\psi_{\vec{k},s}^{(+)} \sim 
{\widehat{\mathcal{S}}}
\left[
\begin{array}{c}
1 \\
\frac{1}{E+M} \vec{\sigma} \cdot \vec{p} 
\end{array} 
\right]
e^{\imath \vec{p}_p \cdot \vec{r}}
\chi_{\frac{1}{2}m_{s}} \Psi _{A-1} ^{J_R \; M_R} \left( \vec{r}_2,
\vec{r}_3, \ldots \vec{r}_A \right)  \; .
\label{eq:glauber}
\end{equation}
In this expression, the subsequent elastic or ``mildly inelastic''
collisions with ``frozen'' spectator nucleons which the ejectile
undergoes, are implemented through the introduction of the following
operator
\begin{displaymath}
{\widehat{\mathcal{S}} \left( \vec{r}, \vec{r}_2,
\vec{r}_3, \ldots \vec{r}_A \right) }
\equiv \prod _ {j=2} ^{A} \left[ 1 - \Gamma \left( \mid \vec{p}_p
\mid, \vec{b} - \vec{b_j}
\right) \theta \left( z - z_j \right) \right] \; ,
\end{displaymath}
where the profile function for elastic $pN$ scattering reads
\begin{eqnarray*}
\Gamma (p_p,\vec{b}) = \frac{{\sigma^{tot}_{pN}}
(1-i {\epsilon_{pN}})}
{4\pi \beta_{pN}^{2}} \: exp \left(
-\frac{b^{2}}{2\beta_{pN}^{2}} \right) \; .
\label{eq:profilefunction}
\end{eqnarray*}
In practice, for a given ejectile's lab momentum the following input is
required : the total proton-proton and proton-neutron cross sections ${\sigma^{tot}_{pN}}$,  
the slope parameters ${\beta_{pN}}$ and the ratios of the real to
imaginary scattering amplitude ${\epsilon_{pN}}$. 
 
\section{Results}

\subsection{$\mathbf{^{16}}$O$\mathbf{(e,e'p)}$ : 
$\mathbf{\omega}$ = 0.439~GeV and $\mathbf{Q^2}$ = 0.8~(GeV/c)$\mathbf{^2}$} 

\begin{figure}
\vspace{-0.8cm}
\begin{center}
\includegraphics[totalheight=11.0cm]{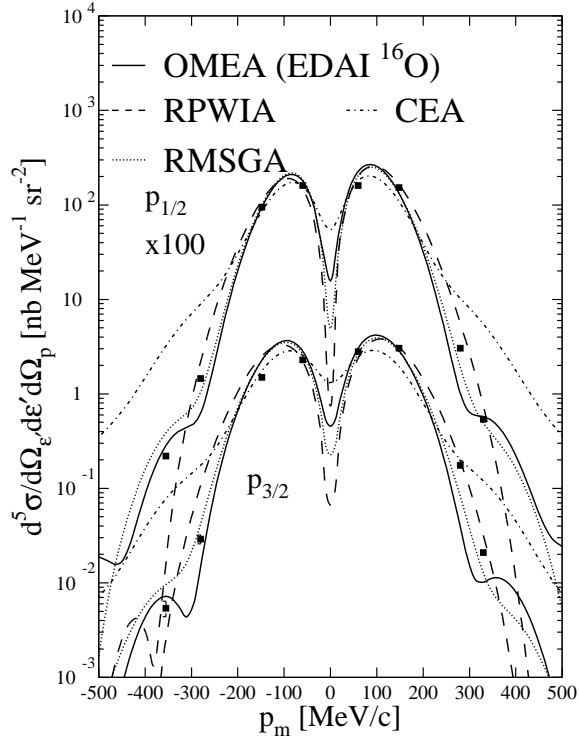}
\end{center}
\vspace{-1.3cm}
\caption{CEA, OMEA, RMSGA and RPWIA calculations (see text for
details) for $^{16}$O$(e,e'p)$
differential cross sections in quasi-perpendicular
kinematics at $\epsilon$ = 2.4~GeV, q = 1~GeV/c and $\omega$ =
0.439~GeV. The calculations use the CC1 off-shell electron-proton
coupling.  The data are from Ref.~\cite{gao}.}
\label{fig:gao}
\end{figure}

We have compared our relativistic calculations to recent quasi-elastic
$^{16}$O$(e,e'p)$ data from JLAB.  In these high-resolution
experiments at $Q^2$ = 0.8~(GeV/c)$^2$ separated structure functions
and polarization transfer data were obtained. In the considered
kinematics the de Broglie wavelength of the ejected proton is of the
order $\lambda _{db} \approx 0.2$~fm and one may expect the
eikonal and Glauber framework to be applicable.  At the same time, at
these kinetic energies, optical potentials are still readily
available. In Figure~\ref{fig:gao} we display the calculated
differential cross sections against the missing momentum.

\begin{figure}
\begin{center}
\includegraphics[totalheight=8.8cm]{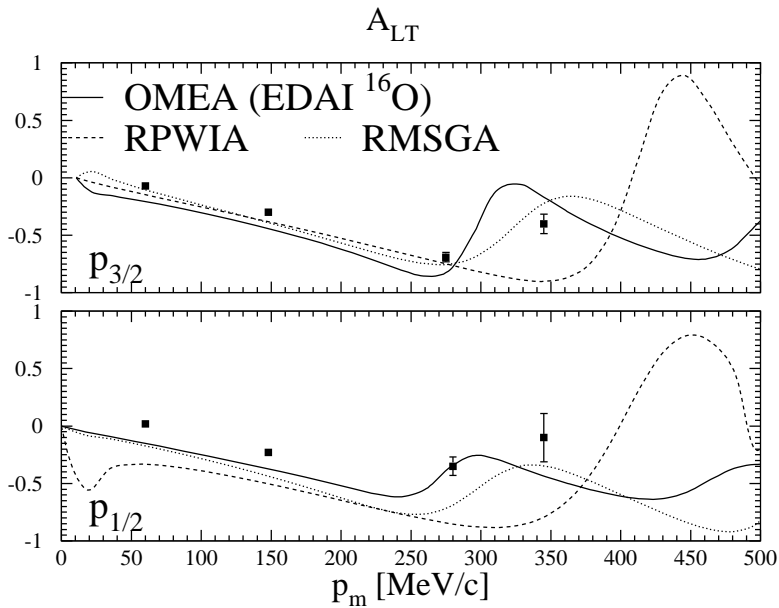}
\vspace{-1cm}
\caption{Left-right asymmetry for the quasi-elastic $^{16}$O$(e,e'p)$
cross sections of Fig. \ref{fig:gao}.}
\label{fig:alt}
\end{center}
\end{figure}

The different curves all use the same bound-state wave
functions and electron-proton coupling but differ in the way the FSI is
treated.  The adopted techniques are \vspace{-0.3cm}
\begin{enumerate}
\item A Relativistic Plane Wave Impulse Approximation (RPWIA)
calculation in which the ejectile is described by a relativistic plane
wave. \vspace{-0.3cm}
\item
A so-called Consistent Eikonal Approximation (CEA) calculation
\cite{dimi1}.  Here, the eikonal phase (\ref{eq:eikonalphase}) is
calculated from the relativistic scalar and vector potentials which
determine also the bound-state wave functions. \vspace{-0.3cm}
\item
An Optical-Model Eikonal Approximation (OMEA) calculation.  Here,
the eikonal phase (\ref{eq:eikonalphase}) is calculated from the
relativistic optical potentials as they are derived from global fits
to elastic proton-nucleus data.  We use the energy-dependent mass
number independent (EDAI) version of Ref.~\cite{cooper}. \vspace{-0.3cm}  
\item
A Relativistic Multiple-Scattering Glauber Approximation (RMSGA)
calculation.  In this approach, the scattering state is calculated on
the basis of Eq.~(\ref{eq:glauber}). \vspace{-0.3cm}
\end{enumerate}
The CEA approach, despite the fact that it obeys orthogonality and
unitarity constraints, fails to predict the missing-momentum
dependence of the measured cross sections. The RMSGA and the OMEA
approach, on the other hand, both seem to give a reasonable
description of the data.  A similar remark holds for the separated
structure functions (not shown here) and the polarization transfer
data.  Several relativistic calculations based on optical potentials
nicely agreed with the $^{16}$O$(e,e'p)$ data set \cite{gao}.  The
novel thing about the results displayed in Figure~\ref{fig:gao}, is
that also an out-of-the box relativistic calculation within the
Glauber framework produces fair results.  To illustrate this further,
we display in Fig.~ \ref{fig:alt} the left-right asymmetry
\begin{equation}
A_{LT} = \frac{\sigma(\phi=0^{\circ}) - \sigma(\phi =
180^{\circ})}{\sigma(\phi=0^{\circ}) + \sigma(\phi = 180^{\circ})} =
\frac{v_{LT}W_{LT}}{v_{L}W_{L}+v_{T}W_{T}+v_{TT}W_{TT}} \; .
\end{equation}
This quantity is highly sensitive to relativistic effects
\cite{gao,jan99}.  At higher missing momenta, the $A_{LT}$ data
strongly deviate from the RPWIA predictions pointing towards important
FSI effects.  Both the Glauber-based RMSGA and the optical-potential
based OMEA approach provide a reasonable description of the high-$p_m$
data points.  

Spectroscopic factors are extracted quantities which normalize a
theory to the data.  Recent investigations \cite{lapikas00,zhalov}
point towards the rather intriguing situation that the spectroscopic
factors extracted at higher values of Q$^2$ turn out to be
substantially larger than those derived at lower values of Q$^2$
\cite{ingo}.  This phenomenon, which has been dubbed as ``quenching
disappearance'' may point towards a scale dependence in nuclear
physics. It turns out, however, that all the ``higher'' values for the
spectroscopic factors are obtained within the Glauber approach whereas
the lower values are extracted by comparing optical-model calculations
with data.  One could wonder whether there is any real physics behind
this.  It may as well be that these observations regarding the value
of spectroscopic factors reflect the inability of the theorists to
treat the FSI in $(e,e'p)$ processes and to provide model calculations
that ensure a smooth transition between the ``low-energy''
(optical-potential) and the ``high-energy'' (Glauber inspired) regime.
Our framework has the feature of being able to deal with optical
potentials as well as the Glauber approach, as long as one works in
kinematic regimes where the eikonal approximation appears plausible.
We asked ourselves the question whether the Glauber and optical
potential approach produce the same spectroscopic factors for the
kinematics of the $^{16}$O results in Figs. \ref{fig:gao} and
\ref{fig:alt}.  Therefore, a fit of the calculations to the
data was performed, including the results for the differential cross
sections and the separated structure functions.  The results are
summarized in Table \ref{tab:spec}.  We find that the Glauber and
optical potential framework lead to almost identical spectroscopic
factors, which are significantly larger than what was typically found
at lower values of Q$^2$.

\begin{table}
\begin{center}
\begin{tabular}{ccccc} \hline 
   & OMEA (CC1) 
   & RMSGA (CC1)
   & OMEA (CC2) 
   & RMSGA (CC2)
\\ \hline \hline
$\frac{1}{2}^-$ (g.s) & 0.79 (0.60) & 0.80 (0.38) & 0.82 (1.04) 
& 0.82 (0.95) \\
$\frac{3}{2}^-$ (6.31~MeV) & 0.96 (1.78) & 0.96
(0.75) & 1.00 (3.48) & 1.00 (2.19) \\
\hline 
\end{tabular}
\end{center}
\caption{The spectroscopic factors (normalized to 1) for the
$^{16}$O($e,e'p$) reaction in the kinematics of
Fig.~\ref{fig:gao}. The obtained $\chi^{2}$ per degree-of-freedom is given
between brackets.}
\label{tab:spec}
\end{table}

\subsection{Nuclear transparency in $\mathbf{^{12}}$C}

\begin{figure}
\begin{center}
\includegraphics[totalheight=7.cm]{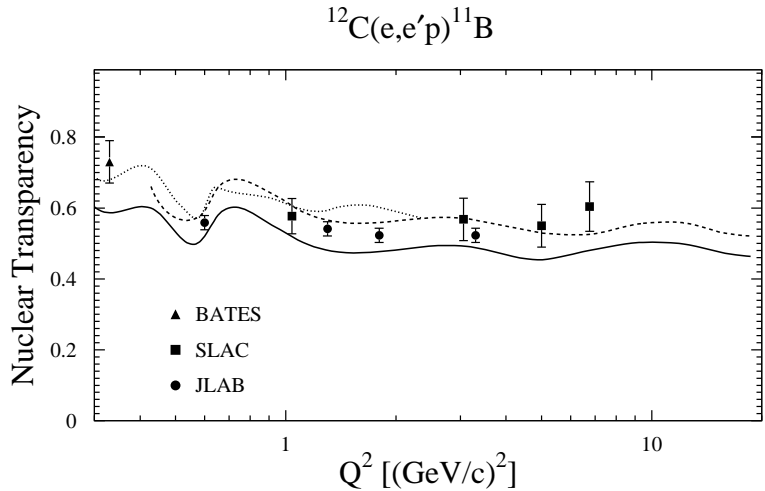}
\vspace{-1cm}
\caption{Nuclear transparency for $^{12}$C$(e,e'p)$ as a function of
Q$^2$.  The theoretical predictions are obtained within the RMSGA
(solid line) and OMEA (dotted line).  The dashed line is the RMSGA
result after including SRC effects. The calculations use the CC2 form
for the electron-proton coupling.}
\label{fig:trans}
\end{center}
\end{figure}

The nuclear transparency provides a measure of the likelihood that a
struck ``nucleon'' escapes from the system.  The $Q^2$ and $A$
dependence of this quantity can provide information about 
point-like configurations (PLC) in the nucleus and the
onset of non-hadronic degrees of freedom.  Experimentally, the
transparency is determined by taking the ratio of the integrated measured
A$(e,e'p)$ strength over certain energy $\Delta E$ and momentum
$\Delta ^{3} k$ ranges 
to the non-relativistic (NR) PWIA prediction.
Typical ranges for light nuclei
are determined by the Fermi momentum  $p_m \le k_F$ and the
excitation-energy where all the single-particle strength is expected
to reside, i.e.\ $E_m \leq 60$~MeV.  We have computed the nuclear
transparency according to
\begin{equation}
T_{th} = \frac {\sum _{\alpha}
\int _0 ^{k_F} d \vec{k} \frac {
\frac {d^5 \sigma}  {d \Omega _p d \epsilon ' d \Omega
_{\epsilon '}} (e,e'p_{\alpha})
                           }
{K \sigma _{ep} ^{CC1} } } 
{\sum _{\alpha} {\int_{0} ^{k_F} d \vec{k} 
\; S_{NRPWIA}^{\alpha}(\vec{k})} } \; ,
\end{equation}
where the sum over $\alpha$ extends over all occupied single-particle
states of the target nucleus.

The description of the nuclear transparency at higher Q$^{2}$ requires
the inclusion of effects from short-range correlations (SRC).  The
inclusion of SRC causes an overall enhancement of the nuclear
transparency with some 10 \%. No Q$^{2}$ dependent increase of the
nuclear transparency with increasing momentum transfers is observed.
As was pointed out in Refs.~\cite{lapikas00} and \cite{zhalov}, the
$^{12}$C nuclear transparencies can be linked to the summed
spectroscopic factor of the 1s and 1p levels.  The Q$^{2} > 1$
(GeV/c)$^{2}$ results in Fig.~\ref{fig:trans} are compatible with a
summed spectroscopic factor which is Q$^{2}$ independent and
substantially larger than what is commonly found at low values of
Q$^{2}$.  Another important finding from Fig.~\ref{fig:trans} is that
when it comes to describing FSI effects in A($e,e'p$) transparency
calculations, the optical-potential and Glauber framework provide
reasonably consistent results, provided that the Glauber calculations
are corrected for SRC effects.  Indeed, in the regime where both the
optical-potential and the Glauber approach appear justified, both
predict comparable transparency results, including the structures in
the Q$^2$ dependence of the transparency at lower values of Q$^2$.

\section{Conclusion}
Summarizing, a flexible framework for
modeling A$(e,e'p)$ processes at high values of the four-momentum
transfer has been presented.  The model is relativistic and fully
unfactorized and can be used in combination with optical potentials,
when available, as well with the Glauber multiple-scattering approach
for the treatment of the final-state interactions.  One of the major
conclusions of our investigations is that the Glauber framework, in
which the final-state interactions are computed directly from the
elementary proton-nucleon processes, works reasonably well even when
it comes to predicting polarization observables and separated
structure functions.  The relativistic effects on the
total cross sections are relatively small.  Some structure functions,
for example the longitudinal-transverse interference term $R_{LT}$,
turn out to be highly sensitive to relativistic effects, an
observation which was earlier made in d$(e,e'p)$ studies.

Our unfactorized and relativistic Glauber framework provides a
reasonable description of the measured $^{12}$C$(e,e'p)$ nuclear
transparencies.  In these transparency calculations, we observe a
relatively ``smooth'' transition between the energy regime where
optical potentials can be used and the high-energy regime where the
use of the Glauber framework appears legitimate.  In the energy regime
where both the ``optical potential'' and the ``Glauber'' treatment of
final-state interactions appear justified, both approaches give
similar results provided that the free proton-nucleon cross sections
are corrected for short-range correlations in the target nucleus.
From our calculations we have two arguments which seem to support the
``quenching disappearance'' phenomenon at higher values of Q$^2$.  The
spectroscopic factors which are extracted from comparing our
relativistic eikonal calculations to the extensive $^{16}$O$(e,e'p)$
data set at Q$^2$ = 0.8~(GeV/c)$^2$, which includes separated
structure functions and differential cross sections, are substantially
higher than the values obtained at lower values of Q$^2$.  One
striking feature is that the Glauber and ``optical potential''
treatment of final-state interactions, which are intrinsically very
different, lead to almost identical values for the extracted
spectroscopic factors.  Also a comparison of the relativistic eikonal
predictions with the available $^{12}$C$(e,e'p)$ transparency data,
is consistent with values of the summed spectroscopic strength which
almost exhaust the sum-rule value when integrated over the first 60
MeV in missing energy of $^{11}$B.

\end{document}